\documentclass[reprint,amsmath,amssymb,onecolumn, pre,longbibliography]{revtex4-1}
\usepackage[pdftex]{graphicx}
\usepackage{bm}
\usepackage{amssymb}
\usepackage{latexsym}
\usepackage{ulem}
\usepackage{color}
\usepackage{amsmath}
\usepackage{lipsum}
\usepackage{floatrow}
\usepackage{epstopdf}

\begin{document}
\title{Wall slip of complex fluids: interfacial friction versus slip length
}
\author{Benjamin Cross}
\affiliation{Univ. Grenoble Alpes, CNRS, LIPhy, F-38000 Grenoble, France.}
\author{Chlo\'e Barraud}
\affiliation{Univ. Grenoble Alpes, CNRS, LIPhy, F-38000 Grenoble, France.}
\author{Cyril Picard}
\affiliation{Univ. Grenoble Alpes, CNRS, LIPhy, F-38000 Grenoble, France.}
\author{Liliane L\'eger}
\affiliation{Laboratoire de Physique des Solides, CNRS, Universit\'e Paris-Sud, Universit\'e Paris-Saclay, 91405 Orsay Cedex, France}
\author{Fr\'ed\'eric Restagno}
\affiliation{Laboratoire de Physique des Solides, CNRS, Universit\'e Paris-Sud, Universit\'e  Paris-Saclay, 91405 Orsay Cedex, France}
\author{\'Elisabeth Charlaix}
\email{Elisabeth.Charlaix@univ-grenoble-alpes.fr}
\affiliation{Univ. Grenoble Alpes, CNRS, LIPhy, F-38000 Grenoble, France.}
\begin{abstract}
Using a dynamic Surface Force Apparatus, we demonstrate that the notion of  slip length  used to describe the boundary flow of simple liquids, is not appropriate for viscoelastic liquids. Rather, the appropriate description lies in the original Navier's  partial slip boundary condition, formulated in terms of an interfacial friction coefficient. We establish an exact analytical expression to extract the interfacial friction coefficient from oscillatory drainage forces between a sphere and a plane, suitable for  dynamic SFA or Atomic Force Microscopy non-contact measurements.  We use this model to investigate the boundary friction of  viscoelastic polymer solutions over 5 decades of film thicknesses and one decade in frequency. The proper use of the original Navier's condition describes accurately the complex hydrodynamic force up to scales of tens of micrometers, with a simple "Newtonian-like" friction coefficient, not frequency dependent, and reflecting closely the dynamics of an interfacial depletion layer at the solution/solid interface.
\end{abstract}
\maketitle

Flow of complex liquids are familiar and useful. Unlike Newtonian fluids, they display complex bulk rheological behavior, non-linear and frequency-dependent. But the way they flow  also involves their boundary conditions on solid surfaces. The boundary condition (b.c.) is relevant for small scale flows, occurring for instance in bio-medical applications, microfluidic devices, food or oil engineering,  but also for the faithful characterization of the  bulk rheology \cite{barnes_review_1995,Denn2001,Reiter2000,Jacobs2010,hatzikiriakos_wall_2012}.

As in the case of simple liquids, the slippage of complex fluids at walls is commonly characterized by a slip length $b$, defined by  the ratio of the  
fluid velocity at the solid surface to the shear rate at the wall: $v_{\rm slip} = b \partial v(z) / \partial z$, with $z$ the direction normal to the boundary. 
But the notion of slip length, now well established and understood in the case of simple fluid flowing on various types of solid surfaces  \cite{Thompsonrobbins90,BarratBocquetPRE94, thompsontroian97,pitPRL2000,netoreview, BocquetSecchiNature2016,Leroy2012} or Newtonian polymer melts \cite{baumchen_slip_2010,henot_comparison_2017}, is far from being obvious in the case  of more complex fluids. 
We show here experimentally that the appropriate quantity to describe the boundary slippage of complex fluid without ambiguity, is indeed not a slip length, but rather a liquid/wall friction coefficient, as originally stated by Navier \cite{Navier}.

We demonstrate this on the particular example of semi-dilute, viscoelastic poly-electrolyte solutions. Water-soluble polyelectrolytes of high molecular weight are commonly use to thicken water solutions at an affordable price, as small concentrations are sufficient to increase significantly the viscosity of the solution \cite{DobryninRubinstein2005}. 
Water-soluble polyelectrolytes of high molecular weight have been reported to  display large slip on various types of solid surfaces \cite{Chauveteau1,Chauveteau2,CayerBarriozTL2008,CuencaPRL2013}.  This large slip has been attributed to the presence of a depletion layer at the solution/solid interface, 
 i.e. a layer with a lower concentration of polymer or even with pure solvant, whose viscosity significantly  lower than that of the solution induces an apparent slip boundary condition \cite{de_gennes_polymer_1981,barnes1995review,Kuhl-Israelachvili1998,HornVinogradovaJCP2000,knoben_direct_2007}. 
Here we use Partially Hydrolyzed PolyAcrylamide (HPAM) semi-dilute solutions, in conditions which were otherwise well characterized by other groups and are industrially  used for Enhanced Oil Recovery or water purification \cite{CuencaPRL2013} (SNF Flopaam 3630S, molecular weight 20.10$^6$~g/mole solved in deionized water at concentration from 0.8 to 1.6~g/l). The boundary flow of these solutions  is studied with a dynamic Surface Force Apparatus (dSFA, \cite{rsi2002, Garcia2016}) by confining them between a pyrex sphere (radius $R=3.3$~mm) and a pyrex plane of very low roughness (2~\AA~rms as measured by AFM).  The set-up covers sphere-plane distances $D$ ranging over 5 orders of magnitudes from $0.1$ nm to 15 $\mu$m, allowing one to bridge the macroscopic flow behavior of the liquid to its interfacial hydrodynamics.

\begin{figure}[hbtp]
 \includegraphics[width=9cm]{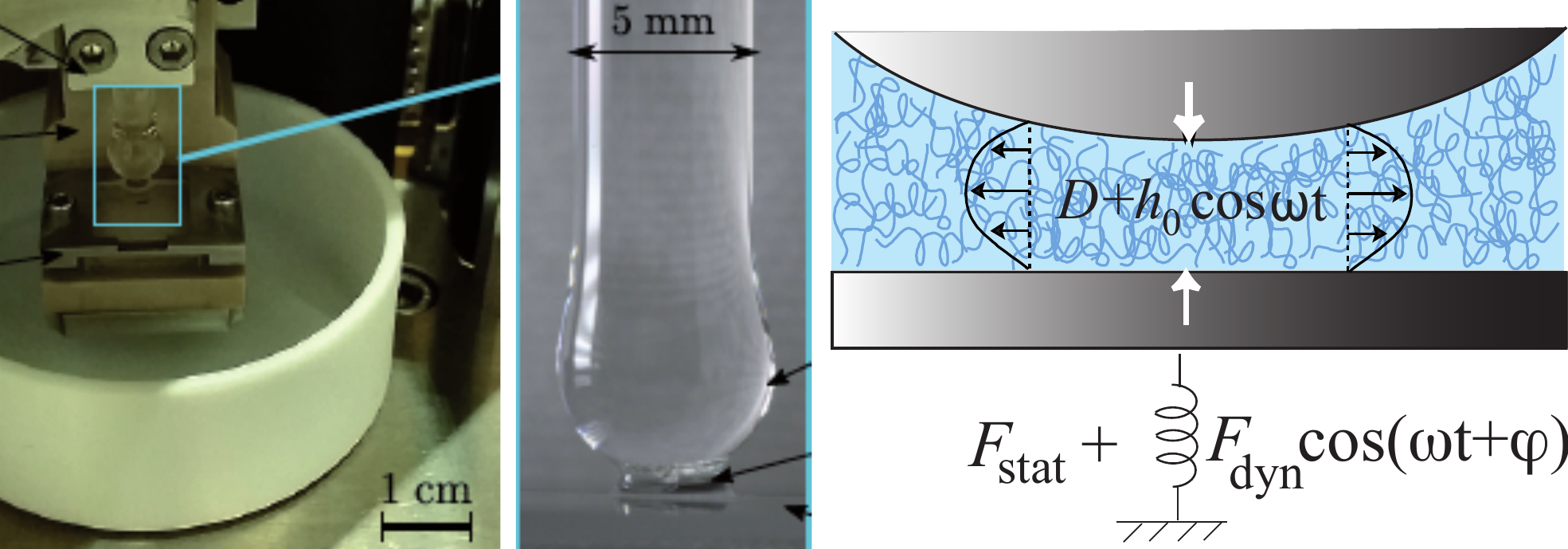}
\caption{\label{fig:1}The dynamic Surface Force Apparatus. Left: large view of the sphere-plane contact. Center: detailed view of the contact. Right: Schematic of the flow.}
\end{figure}

The sphere is driven normally to the plane at a low drift velocity ($\dot D/D < 10^{-2}$ s$^{-1}$). An oscillatory motion of small amplitude ($h_0/D < 10^{-2}$)  at angular frequency $\omega$ is added to the slow drift motion. The relative sphere-plane displacement as well as the force acting on the plane, are measured by two independent external interferometric sensors. From these measurements we get the steady-state sphere-plane distance $D$  and  interaction force $F_{\rm stat}$, the dynamic amplitude $h_0$ of the sphere-plane oscillatory displacement which is chosen as the phase origin, the dynamic amplitude $F_{\rm dyn}$ and phase-shift $\varphi$ of the oscillatory  interaction force at the frequency $\omega$, and finally, the   linear force response or mechanical impedance, defined as: 
\begin{equation}
\tilde Z (D,\omega) = \frac{F_{\rm dyn}e^{i\varphi}}{h_0}=Z_R+iZ_I
\label{Z}
\end{equation}

\begin{figure}[htbp]
  \includegraphics[width=9cm]{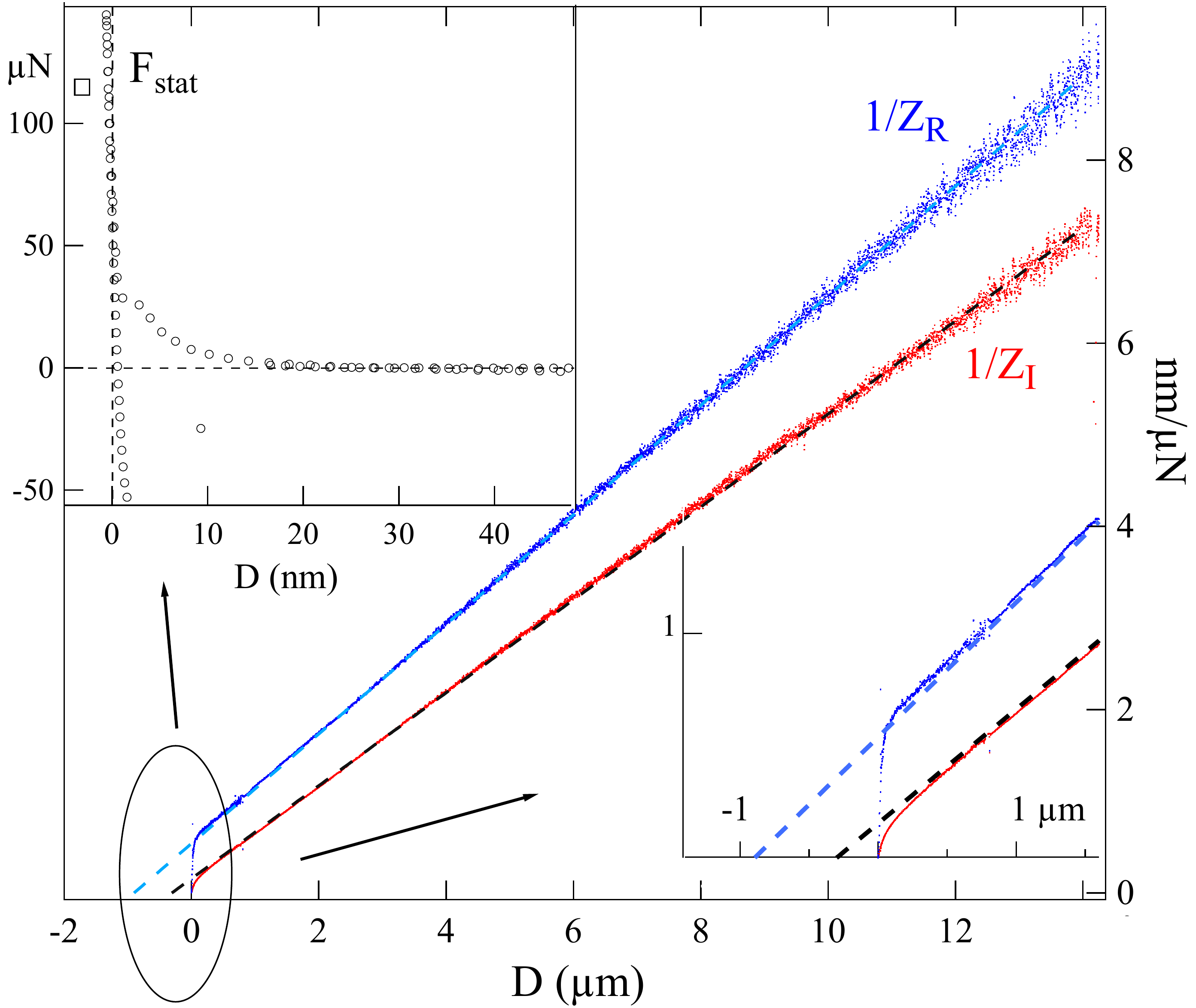}
  \hskip 3mm
   \includegraphics[width=7.8cm]{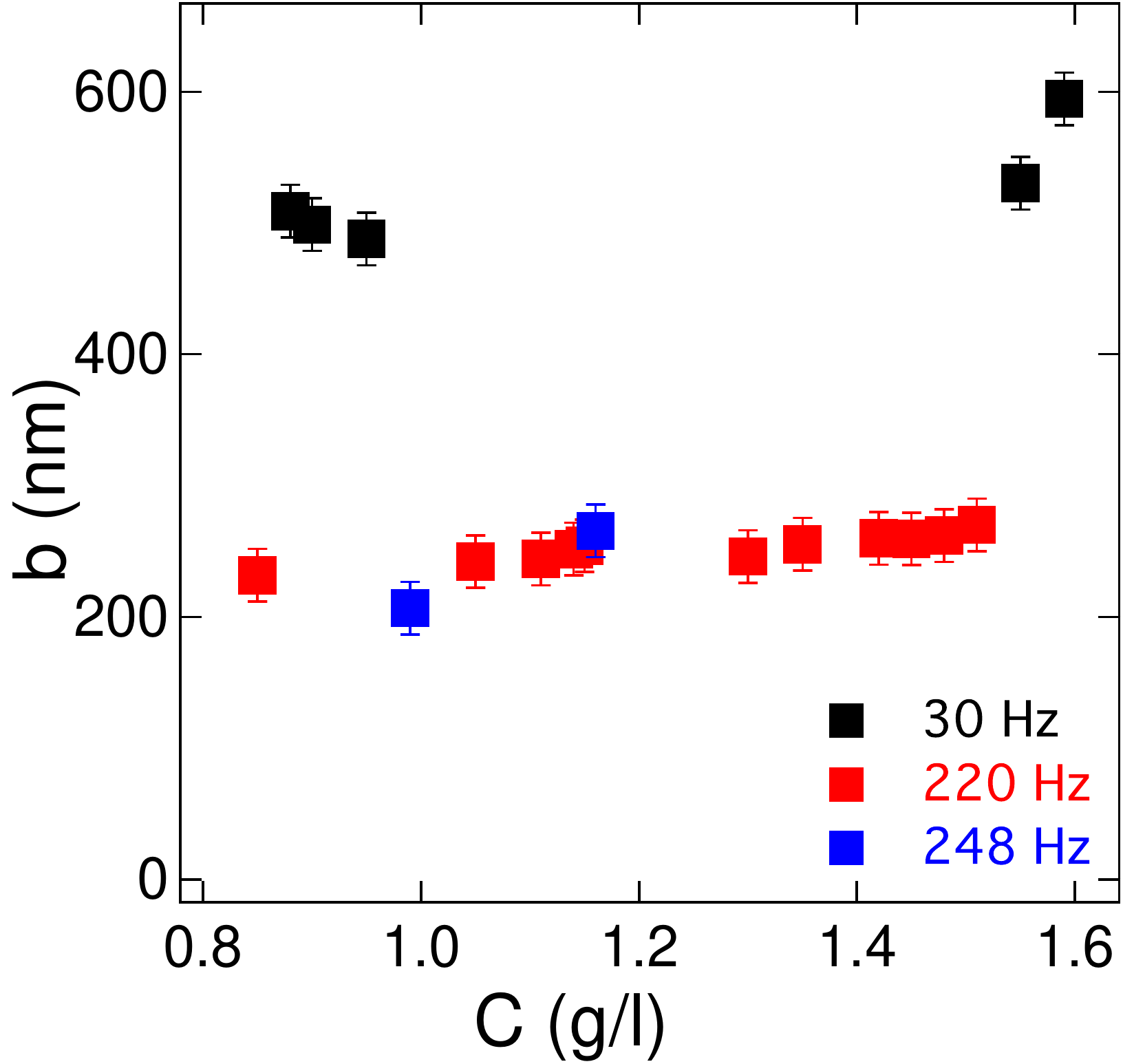}
\caption{Left: inverse of the components of the dynamic force response $\tilde Z$ measured in an HPAM solution at a frequency of 220 Hz, as a function of the sphere-plane nominal distance $D$. Blue: $1/Z_R$, red: $1/Z_I$. The top-left inset shows the quasi-static interaction force  $F_{\rm stat}$, whose jump-to-contact defines the origin of distances. The down-right inset is an enlargement  at the submicrometric scale. Right: slip length defined from the extrapolation length of $1/Z_I$  (intercept with the x-axis of the black dashed line of the left plot), as a function of the concentration of the solution, for various frequencies.   }
\label{fig:2}
\end{figure}

For a viscoelastic liquid of complex shear modulus $\tilde G=G_R+{\rm i} G_I= {\rm i}\omega (\eta_R -{\rm i} \eta_I)$,  the hydrodynamic force response in this oscillatory drainage flow, in the case of a no-slip boundary conditions at walls, is \cite{Pelletier94b}:
\begin{equation}
\tilde Z (D,\omega) = \frac{6 \pi R^2 \tilde G}{D}= \frac{6 \pi R^2 {\rm i} \omega\tilde \eta}{D}
\label{Pelletier}
\end{equation}
Therefore it is convenient to characterize the bulk viscoelasticity of the solutions by plotting 1/$Z_{\rm R}$ and 1/$Z_{\rm I}$ as a function of the sphere-plane distance, looking at large values of $D$ (fig.\ref{fig:2}). At large distance a well-defined linear behavior is observed,  demonstrating the bulk viscoelastic character of the solutions. From the  slope of $1/Z_R$ and  $1/Z_I$ we extract the complex shear modulus components $G_R$ and $G_I$.  
 
However, contrary to the prediction of eq. \ref{Pelletier}, the extrapolated far-field linear dependency of $Z_R^{-1}$ and $Z_I^{-1}$ does not point toward the distance origin, but towards some negative values of $D$. This is usually the signature of a slippage effect at the solid-liquid interface. More specifically, it is known in simple fluids that if a slip length $b$ defines the slip boundary condition on each surface then the hydrodynamic force at large distance $D \gg b$ involves the "hydrodynamic thickness"  $D+2b$ instead of the actual distance $D$. 
It is thus tempting to describe the measurements by deriving a  slip length from the extrapolation lengths of $1/Z_R$ and  $1/Z_I$. However when doing so, two difficulties  appear. First, the two linear extrapolations 
actually point towards two different origins, which is not consistent with a single, well-defined slip length. Instead, the extrapolations tend to show that the slip length is complex, possessing a real and an imaginary components. Second, if one determines the slip length as in simple liquids from the extrapolation length of the damping $1/Z_I$,  one finds that it depends significantly on the frequency of the oscillatory flow (see fig. \ref{fig:2}). At a concentration $\sim$ 1 g/l for instance, the slip length decreases by a factor larger than 3 when the frequency increases from 30 Hz to 220 Hz.  This complex and  frequency-dependent  behavior of the slip length does not reflect  the mechanism usually producing large slip of polymer solutions on solid surfaces, which involves the presence of an interfacial  depletion layer made of pure solvant \cite{Kuhl-Israelachvili1998,HornVinogradovaJCP2000}: in the presence of a purely Newtonian lubricating layer, one should expect a purely dissipative and fully Newtonian friction mechanism of the polymer solution onto the solid surface.

These two difficulties arise  because the slip length $b$ is actually defined as the ratio $\eta/\lambda$ of the  liquid viscosity to the interfacial friction coefficient first introduced by  Navier's in its original statement of the boundary condition \cite{Navier}:
\begin{equation}
\lambda v_{\rm slip} =  \eta \frac{\partial v(z)}{\partial z}
\label{Navier}
\end{equation}
For Newtonian fluids the viscosity is a constant quantity, not dependant on the frequency or shear rate,
and the slip length thus provides a convenient image of the interfacial friction. However the mixture of bulk and  interfacial properties entering in the slip length definition raises ambiguities when the fluid is  non-Newtonian.  

\medskip

Keeping this in mind,  we proceed to extract directly the interfacial friction coefficient from the force measurements in  oscillatory drainage flow experiments. For this purpose, we calculate the hydrodynamic force exerted by the fluid drainage   between  the sphere  and the plane, when the viscoelastic fluid undergoes the Navier's partial slip boundary condition eq. (\ref{Navier}) on both surfaces. 
We restrict to the linear response limit $h_0\ll D$, in which  all time variations are harmonic  at the forcing frequency $\omega$,    and all time-varying quantities  are characterized by 
their complex amplitude. In these conditions the (amplitude of the) stress tensor in the liquid is 
$ \overline{\overline \sigma} =\tilde \eta (\nabla \vec{ \tilde v} + ^T \nabla \vec{ \tilde v})  - \delta \tilde P \overline{\overline I}$, where $\delta \tilde P$ is the amplitude of the dynamic pressure
inducing the hydrodynamic force
\begin{equation}
  \tilde F_\text{dyn} = \int_0^\infty 2\pi r  \delta \tilde P(r) dr
\end{equation}
The Navier's boundary condition (\ref{Navier}) is used with  a complex friction coefficient $\tilde \lambda$. In a Maxwell model of the interface, $1/\tilde \lambda=1/\lambda_R+i/\omega k$, where $k$ is the interface stiffness and $\lambda_R$ the dissipative friction coefficient.

\begin{figure}[t]
\includegraphics[width=5 cm]{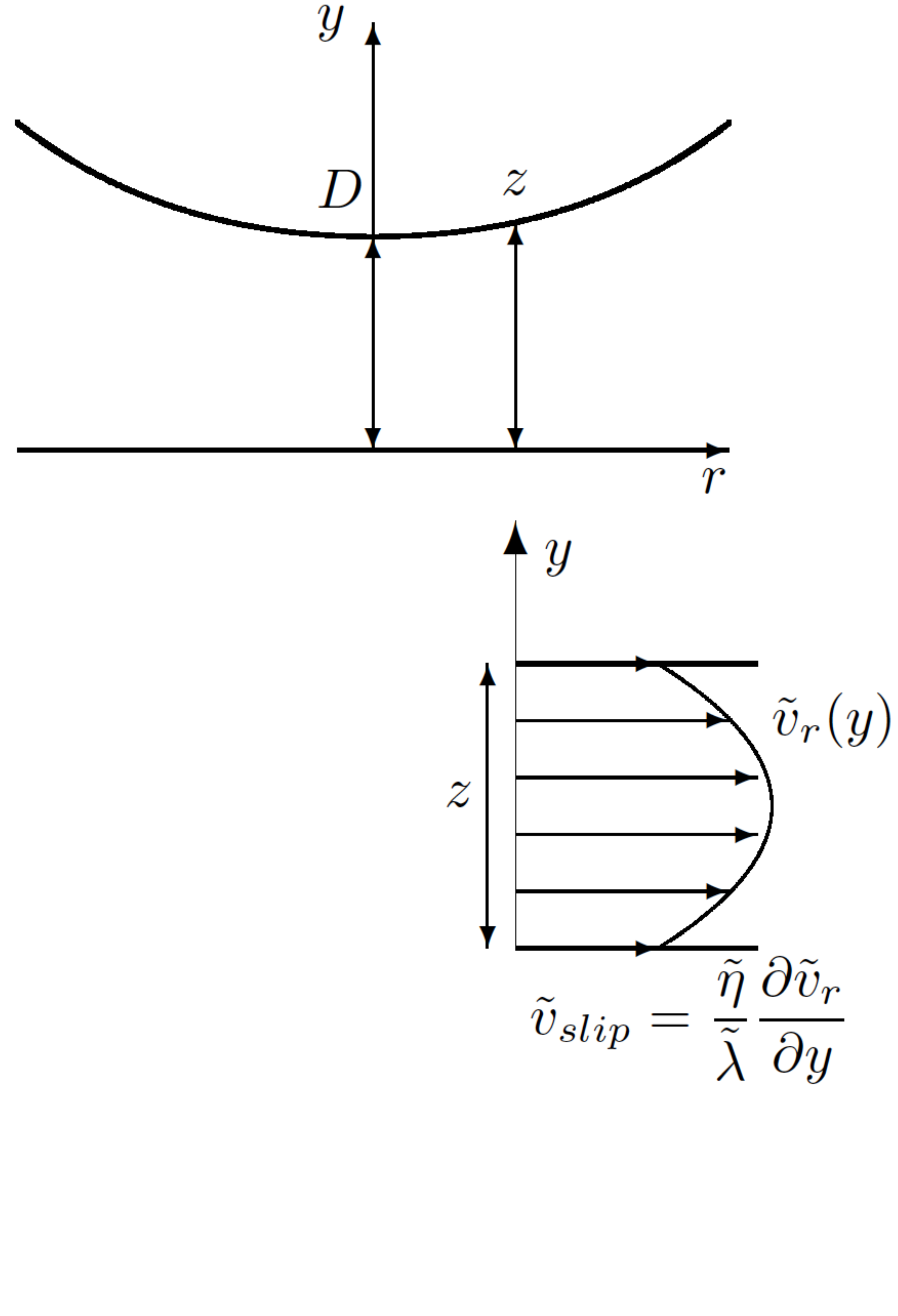}\hfil\includegraphics[width=9 cm]{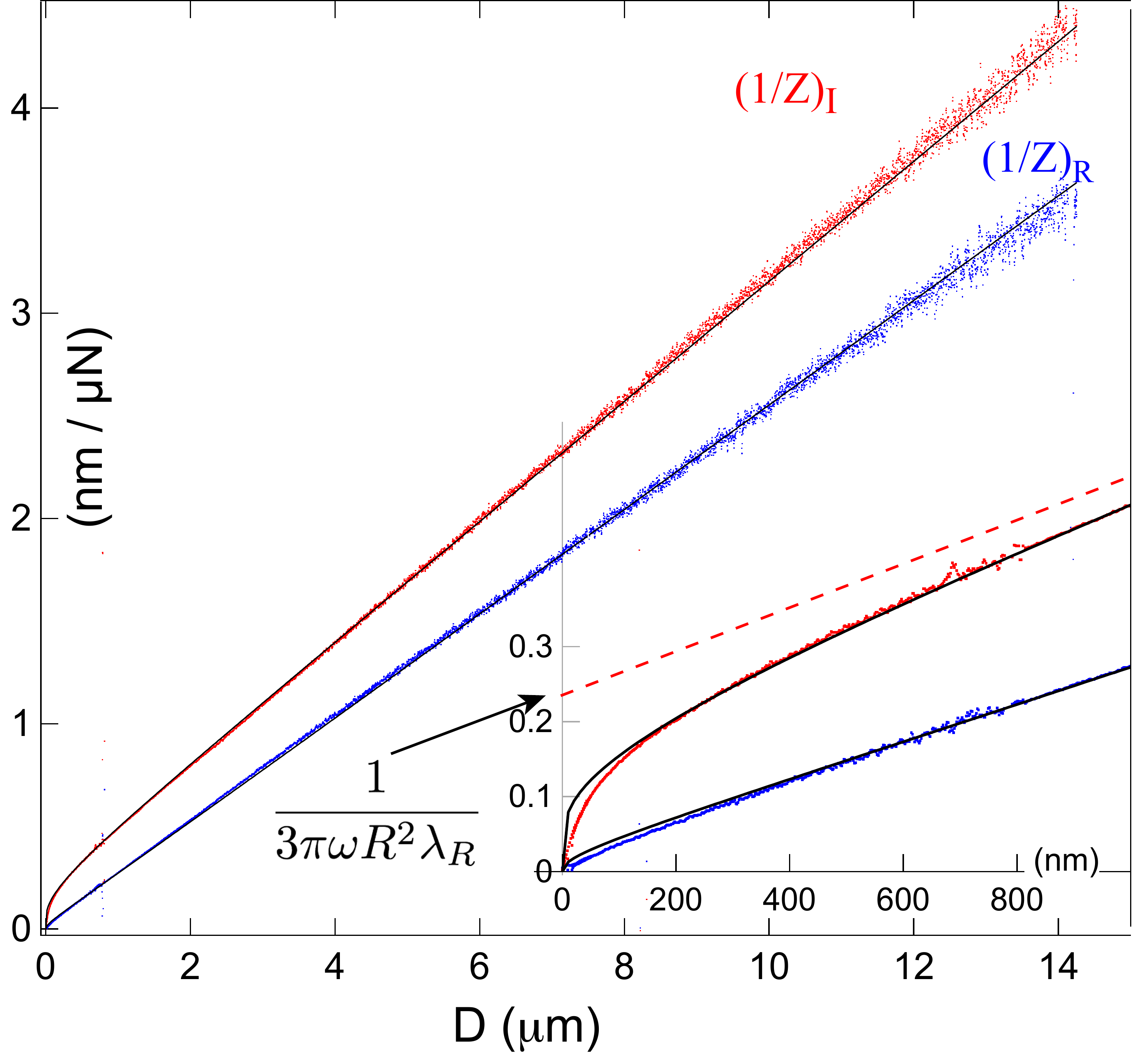}
\caption{\label{fig:3}Left: schematic of the sphere-plane geometry and of the flow profile. Right: elastic component ($1/\tilde Z)_R$ (blue dots) and dissipative component ($1/\tilde Z)_I$ (red dots)  of the inverse of the  force response measured in an HPAM solution at a frequency of 220 Hz. The black continuous lines are the components of the theoretical expression (\ref{Hocking_new}) fitted with a real-valued boundary friction coefficient $\tilde \lambda=\lambda_R$.  Inset: enlargement below the micrometric scale. 
The red dashed line is the extrapolation of the far-field behavior of $(1/\tilde Z)_I$. }
\end{figure}

At small distance $D\ll R$, most of the hydrodynamic force originates from regions where the two solid surfaces are almost parallel. In these conditions, for angular frequencies such that $T=2\pi /\omega \gg \vert \tilde \eta \vert /\rho RD$ inertia is negligible, and the average flow velocity is given by the lubrication approximation \cite{Hocking1973}:
\begin{equation}
\tilde u(r) =\frac{1}{z}\int_0^z \tilde v_r(r,y)dy = - \frac{1}{12\tilde \eta} \frac{d\delta \tilde P}{dr}\left (z^2+6z\frac{\tilde \eta}{\tilde \lambda} \right )
\label{lubrication}
\end{equation}

where $z(r)$ is the nominal gap between the surfaces at distance $r$ from the axis. Note that the fluid velocity and the dynamic pressure $\delta \tilde P$ are of first order in $h_0$, so that only the nominal gap $z(r)$, equal to  $z=D+r^2/2R$ in the parabolic approximation,  enters in (\ref{lubrication}). 
For rigid solid surfaces, the average velocity  $\tilde u(r)$ obeys the conservation relation:
\begin{equation}
\frac{d\left(2\pi r z(r) \tilde u(r)\right)}{dr} =- 2\pi r\frac{\partial z(r)}{\partial t}= -2 \pi r{\rm i} \omega h_0 
\label{conservation}
\end{equation}
Eq. (\ref{lubrication}) and (\ref{conservation}) give the following equation for the  pressure: 
\begin{equation}
\frac{d}{dr}\left [ r z^2\left (z+\frac{6\tilde \eta}{\tilde \lambda}\right )\frac{d\delta \tilde P}{dr}  \right ]  = 12{\rm i} \omega r\tilde \eta h_0
\label{pressure}
\end{equation}
which is integrated twice with $rdr=Rdz$ to obtain the hydrodynamic force:
  \begin{eqnarray}
\tilde Z =\frac{\tilde F_\text{dyn}}{h_0}=  \frac{6 \pi R^2 {\rm i}\omega \tilde{\eta}}{D}f^*\left ( \frac{\tilde \eta}{ \tilde \lambda D} \right )
\label{Hocking_new}\\
f^*(\tilde y) = \frac{1}{3\tilde y} \left [\left ( 1+\frac{1}{6\tilde y}\right ) \ln  ( 1+6\tilde y) -1\right ]
\label{fstar}
\end{eqnarray}
Note that the logarithm entering eq. (\ref{fstar}) should be calculated taking into account the complex character of its argument, by ln($re^{i\theta})=$ln$r+$i$\theta$.

\medskip 

Equations (\ref{Hocking_new}) and (\ref{fstar}) generalize the Hocking expression \cite{Hocking1973} derived for  a Newtonian liquid slipping on the solid surfaces with a slip length $b$. In the Hocking expression, the factor $f^*$ has the same mathematical expression as in eq. (\ref{fstar}), but it depends only on the  simple ratio $D/b$. In the present non-Newtonian case,  we see that a "complex slip length" $\tilde b = \tilde \eta/\tilde \lambda$ governs the hydrodynamic force, which explains the two extrapolation lengths observed in figure (\ref{fig:2}). The complex character of the slip length reflects the phase difference between the boundary slip velocity and the bulk velocity, and the ratio $\tilde b/D$ reflects the impact in amplitude and phase of the wall slippage in the bulk flow. We can see that for a viscoelastic liquid the slip length $\tilde b$ is complex even if the interfacial friction is purely dissipative, which suggests that the complex frequency-dependent slip length observed above might be only an artifact due to the bulk behavior of the solutions.

\begin{figure}[t]
  \includegraphics[width=8.2cm]{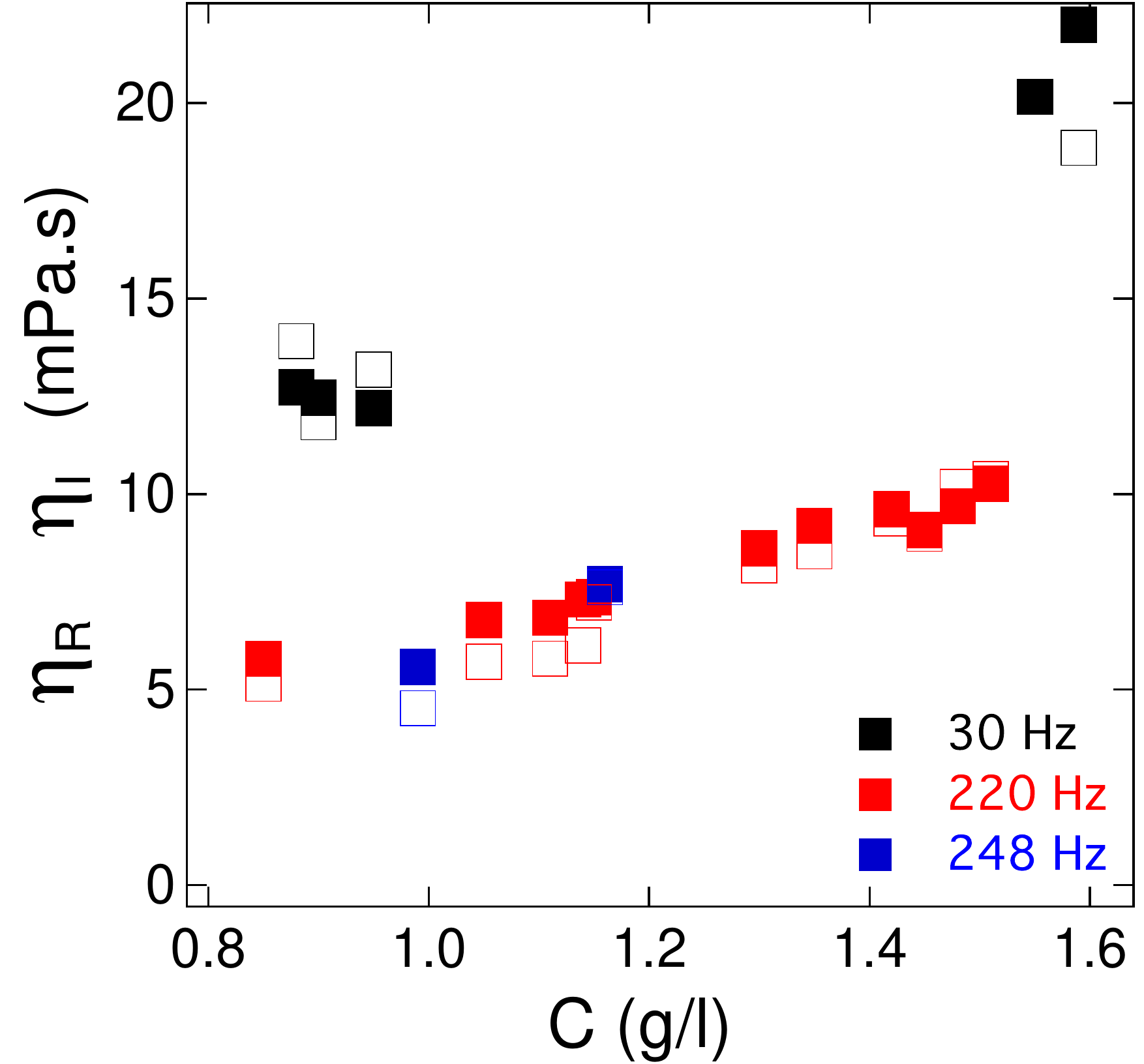}
   \includegraphics[width=8.4cm]{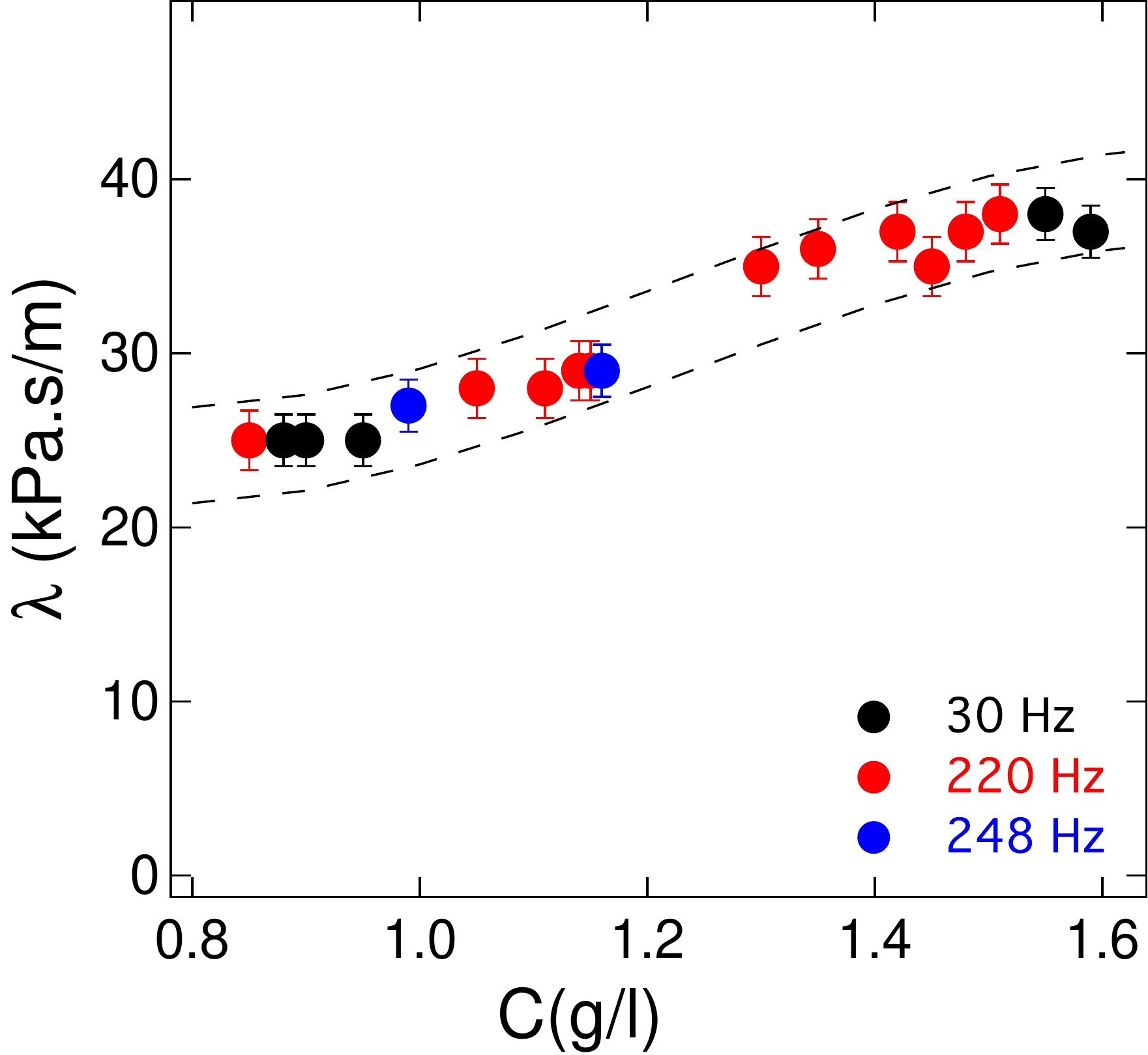}
\caption{Left: components $\eta_R$ ($\blacksquare$) and $\eta_I$ ($\square$) of the visco-elastic modulus $\tilde \eta$ of the solutions as a function of polymer concentration,  at  30 Hz (black), 220 Hz (red), and 248 Hz (blue). Right: interfacial friction coefficient $\tilde \lambda = \lambda_R$ at the solution/solid boundary as a function of the HPAM concentration, at 30Hz ({\Large${\bullet}$}), 220Hz (\textcolor{red}{{\Large$\bullet$}}), and 248 Hz (\textcolor{blue}{{\Large$\bullet$}}). The dashed lines are a guide for the eye.   }  
\label{fig:4}    
\end{figure}

In order to disentangle experimentally the interfacial boundary condition from the bulk properties of the solutions we notice that at large distance $D \gg \vert \tilde \eta/ \tilde \lambda \vert $, the above expression (\ref{Hocking_new})   expands as:
\begin{equation}
\frac{1}{\tilde Z} = \frac{D}{6 \pi R^2 {\rm i}\omega \tilde{\eta}}+ \frac{1}{3 \pi {\rm i} \omega R^2 \tilde \lambda}
\label{farfield}
\end{equation}
Thus, rather than $Z_R^{-1}$ and $Z_I^{-1}$,  the  dynamic quantity providing an independent access  to the   interfacial rheology   is $1/\tilde Z$. More specifically, the boundary friction and stiffness are obtained from the value at origin (intercept on the y-axis) of the linear extrapolation of the far-field components $(1/\tilde Z)_R$ and $(1/\tilde Z)_I$.
Accordingly, the components of $1/\tilde Z$ in our HPAM solutions are plotted in fig.\ref{fig:3}. The linear extrapolation of  $(1/\tilde Z)_R$ points towards the origin within the experimental resolution. This shows that elastic effects in the interface response are negligible: $k\simeq 0$ and $\tilde \lambda$ reduces to $\lambda_R$. The purely dissipative nature of the interfacial friction coefficient reflects the physical mechanism inducing the apparent slip, i.e. the lubrication effect of a Newtonian liquid layer at the boundary. 

The linear extrapolation of $(1/\tilde Z)_I$ (dashed red line fig.\ref{fig:3})  does not extrapolate to zero and gives a first estimation of the liquid/solid friction coefficient $\lambda_R$, in the range of 30 $\mu$Pa.s/nm. For a precise determination of $\lambda_R$ it is however important to compare the whole data to the theoretical expression (\ref{Hocking_new}, \ref{fstar}), because the asymptotic linear dependency of $1/\tilde Z$ with $D$ is  reached only at very large distances $D\gg \vert \tilde \eta \vert / \vert \lambda\vert$. At smaller distances, $(1/\tilde Z)_I$ curves down continuously, converging finally towards the physical origin. 
 This tendency is in excellent agreement with our theory   which provides a very accurate prediction of   both components of the measured dynamic force (see figure \ref{fig:3}). 
\smallskip

Precise estimations of the boundary friction coefficient $\lambda_R$ are obtained   for various concentrations and frequencies by fitting the data to eq.(\ref{Hocking_new} \ref{fstar}), and are plotted in figure \ref{fig:4}.  Unlike the slip length (fig. \ref{fig:2}) the friction coefficient is found essentially insensitive to the frequency over the range studied, which extends on one decade. The absence of  frequency evolution is the signature of a fully Newtonian interfacial friction, as expected for  a slippage mechanism involving a fully depleted, pure water layer at the solution/solid interface. In contrast, the viscoelastic modulii of the solutions vary significantly   in the range of frequencies studied (see figure 4 left). This variation accounts fully for the frequency-dependance of the slip length measured in figure \ref{fig:2}, which is thus due to the bulk rheology of the solutions, and not to the boundary hydrodynamics. Assuming that the viscosity of the depletion layer is that of pure water, its estimated thickness $e_s\simeq \eta_{water}/\lambda$ \cite{vinogradova_drainage_1995,tuinier_polymer_2005} varies between 40 nm to 26 nm for a bulk concentration of polymer  between 0.8 to 1.6 g/l. Furthermore, we should emphasize that in dynamic SFA experiments, the wall shear rate is not spatially uniform, and its range of values changes when the nominal distance $D$ is varied. Considering the excellent agreement of our analytical expression with the data, we can state that the boundary friction coefficient is insensitive to the flow geometry, frequency, and shear rate at wall in the range of $10^{-5}$ to $10^{-2}$s$^{-1}$ probed in the experiment. This shows that the boundary layer inducing the apparent slip is a purely equilibrium layer, whose properties are independent on the applied flow in the range of the above frequencies and shear rates. Such a conclusion cannot be obtained from the properties of the slip length alone, which provides a further proof that the liquid/solid friction coefficient is more appropriate than the slip length to characterize the boundary slip of a complex fluid.

\begin{figure}[t]
  \includegraphics[width=8.4cm]{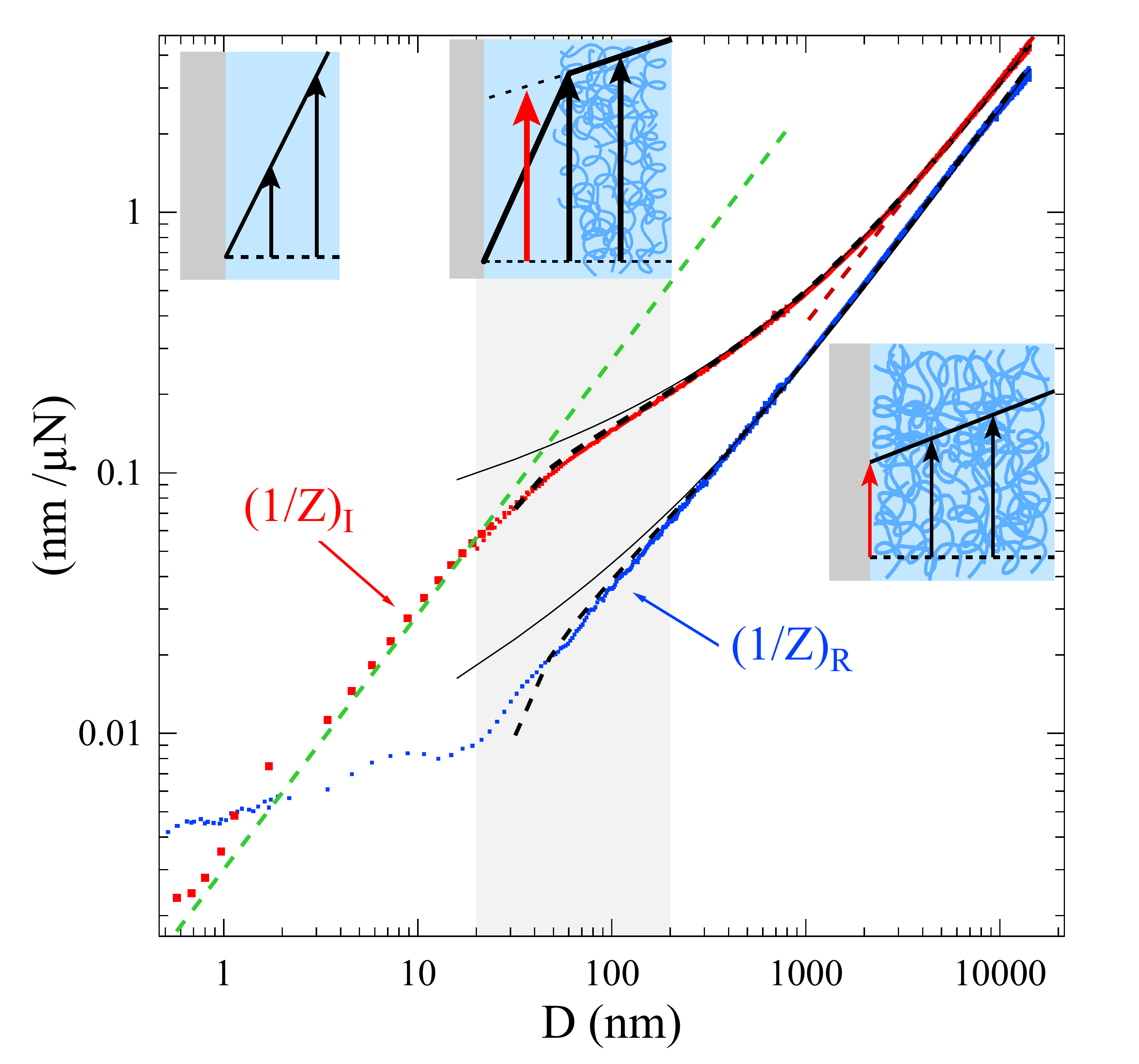}
\caption{Log-log plot of the components of $1/\tilde Z$ (blue dots: real, red dots: imaginary) in a solution of 1.4 g/l at 220Hz, showing the macro-micro transition. The dashed red (resp. green) line corresponds to a Newtonian fluid (resp. water) with a no-slip b.c. at the solid wall. The black continuous lines are the components of  the theoretical expression (\ref{Hocking_new}) taking into account the bulk viscoelasticity and a Navier's b.c. with a real-valued friction coefficient $\lambda_R$. The dashed continuous lines are eq. (\ref{Hocking_new}) with the boundary condition applied inside the liquid, at a distance $e_s/2=\eta_{\rm water}/2\lambda_R=14$ nm of each solid surface. 
}
\label{fig:5}       
\end{figure}

\smallskip

Finally, we discuss the limitations at the microscopic scale of the  model of apparent slip boundary condition. 
The connection of the macroscopic scale to the microscopic scale is shown in figure \ref{fig:5}.
Three areas appear on this figure. Below 20nm, the depletion layer at the solution/solid interfaces is evidenced by the  behavior of the damping ($1/\tilde Z)_I$: 
the latter cannot be distinguished from  a Newtonian liquid of viscosity 0.85 mPa.s,  essentially equal to that of water at the experiment temperature (28 $^o$C), flowing with a no-slip boundary condition on the solid surfaces
(green dashed line in figure \ref{fig:5}).
The microscopic scale of 20 nm is indeed in good agreement with the estimated thickness $e_s\simeq$ 28 nm of the depletion layer 
at this polymer concentration of 1.4 g/l. 
 Thus at distances $D$ smaller than $ e_s$, the sphere-plane gap
 is largely  filled with pure water and the hydrodynamic force resumes to the Reynolds force with a no-slip boundary condition located onto the solid surfaces.  Second, at intermediate distances $e_s < D <10 \ e_s$, the polymer solution flows on a lubricating  layer whose thickness cannot be neglected. Accordingly, as shown theoretically \cite{BarratBocquetPRE94},  the apparent b.c. has to be properly described by 2 independent parameters: the location of the b.c. and the interfacial friction coefficient. For a sharp and large viscosity gradient the appropriate apparent b.c. location lies close to the middle of the depletion layer $e_s$ as sketched in fig. (\ref{fig:5}). Taking this location for the b.c. instead of the actual position of the solid surfaces, amounts to replace $D$ by $D+e_s$ in equations~(\ref{Hocking_new},  \ref{fstar}). We find that it provides a significantly better agreement with the data (dashed black lines in figure 5). Third, at large distances $D> 10 \ e_s$, the finite thickness of the depletion layer can be neglected and one retrieves the apparent slip boundary condition characterized by a  friction coefficient $\tilde \lambda \approx \lambda_R = \eta_{\rm water}/e_s$.

\medskip

In conclusion, we have shown that the slip behavior of a complex fluid at a solid boundary should be consistently analyzed using a boundary friction coefficient rather than a slip length, in order to account faithfully for interfacial rheology effects. In the studied case, the friction coefficient is real-valued and fully Newtonian while the slip length $\tilde \eta /\lambda$  is complex and frequency dependent,  due to the complex bulk rheology of the solutions. This fully Newtonian friction reflects a slip mechanism due to a depletion layer made of pure solvent at the solid-polymer solution interface. Our experiments and analysis show that, beyond a proper description of the boundary condition, the friction coefficient is the relevant quantity for understanding the physical mechanisms governing the interfacial dynamics. 
From a fundamental point of view, the coupling between the generalized lubrication theory developed here and  the unique capabilities of the SFA, opens the route to develop a complete surface rheology for the solid-liquid interface that could be equivalent to what has been recently developed for the fluid-fluid interface \cite{Brenner2013}. In particular it would be interesting to perform experiments on interfaces specifically tailored at the molecular level to provide a non-Newtonian surface friction. Polymer systems are choice candidates to do so.

\bigskip
This research was supported by the ANR program ANR-15-CE06-0005-02.


\begin{thebibliography}{32}%
\makeatletter
\providecommand \@ifxundefined [1]{%
 \@ifx{#1\undefined}
}%
\providecommand \@ifnum [1]{%
 \ifnum #1\expandafter \@firstoftwo
 \else \expandafter \@secondoftwo
 \fi
}%
\providecommand \@ifx [1]{%
 \ifx #1\expandafter \@firstoftwo
 \else \expandafter \@secondoftwo
 \fi
}%
\providecommand \natexlab [1]{#1}%
\providecommand \enquote  [1]{``#1''}%
\providecommand \bibnamefont  [1]{#1}%
\providecommand \bibfnamefont [1]{#1}%
\providecommand \citenamefont [1]{#1}%
\providecommand \href@noop [0]{\@secondoftwo}%
\providecommand \href [0]{\begingroup \@sanitize@url \@href}%
\providecommand \@href[1]{\@@startlink{#1}\@@href}%
\providecommand \@@href[1]{\endgroup#1\@@endlink}%
\providecommand \@sanitize@url [0]{\catcode `\\12\catcode `\$12\catcode
  `\&12\catcode `\#12\catcode `\^12\catcode `\_12\catcode `\%12\relax}%
\providecommand \@@startlink[1]{}%
\providecommand \@@endlink[0]{}%
\providecommand \url  [0]{\begingroup\@sanitize@url \@url }%
\providecommand \@url [1]{\endgroup\@href {#1}{\urlprefix }}%
\providecommand \urlprefix  [0]{URL }%
\providecommand \Eprint [0]{\href }%
\providecommand \doibase [0]{http://dx.doi.org/}%
\providecommand \selectlanguage [0]{\@gobble}%
\providecommand \bibinfo  [0]{\@secondoftwo}%
\providecommand \bibfield  [0]{\@secondoftwo}%
\providecommand \translation [1]{[#1]}%
\providecommand \BibitemOpen [0]{}%
\providecommand \bibitemStop [0]{}%
\providecommand \bibitemNoStop [0]{.\EOS\space}%
\providecommand \EOS [0]{\spacefactor3000\relax}%
\providecommand \BibitemShut  [1]{\csname bibitem#1\endcsname}%
\let\auto@bib@innerbib\@empty
\bibitem [{\citenamefont {Barnes}(1995{\natexlab{a}})}]{barnes_review_1995}%
  \BibitemOpen
  \bibfield  {author} {\bibinfo {author} {\bibfnamefont {Howard~A.}\
  \bibnamefont {Barnes}},\ }\bibfield  {title} {\enquote {\bibinfo {title} {A
  review of the slip (wall depletion) of polymer solutions, emulsions and
  particle suspensions in viscositmer: its cause, character, and curve},}\
  }\href
  {https://ac-els-cdn-com.inp.bib.cnrs.fr/037702579401282M/1-s2.0-037702579401282M-main.pdf?_tid=dbdc6150-ba33-11e7-98fd-00000aab0f6b&acdnat=1509011935_d6b3c0e968b340146c43e42de2d4c8cd}
  {\bibfield  {journal} {\bibinfo  {journal} {J. Non-Newtonian Fluid Mech.}\
  }\textbf {\bibinfo {volume} {56}},\ \bibinfo {pages} {221--251} (\bibinfo
  {year} {1995}{\natexlab{a}})}\BibitemShut {NoStop}%
\bibitem [{\citenamefont {Denn}(2001)}]{Denn2001}%
  \BibitemOpen
  \bibfield  {author} {\bibinfo {author} {\bibfnamefont {M.M.}\ \bibnamefont
  {Denn}},\ }\bibfield  {title} {\enquote {\bibinfo {title} {Extrusion
  instability and wall slip},}\ }\href@noop {} {\bibfield  {journal} {\bibinfo
  {journal} {Annu. Rev. Fluid Mech.}\ }\textbf {\bibinfo {volume} {33}},\
  \bibinfo {pages} {265--287} (\bibinfo {year} {2001})}\BibitemShut {NoStop}%
\bibitem [{\citenamefont {Reiter}\ and\ \citenamefont
  {Khanna}(2000)}]{Reiter2000}%
  \BibitemOpen
  \bibfield  {author} {\bibinfo {author} {\bibfnamefont {G.}~\bibnamefont
  {Reiter}}\ and\ \bibinfo {author} {\bibfnamefont {R.}~\bibnamefont
  {Khanna}},\ }\bibfield  {title} {\enquote {\bibinfo {title} {Real-time
  determination of the slippage length in autophobic polymer dewetting},}\
  }\href@noop {} {\bibfield  {journal} {\bibinfo  {journal} {Phys. Rev.
  Letter}\ }\textbf {\bibinfo {volume} {85}},\ \bibinfo {pages} {2753}
  (\bibinfo {year} {2000})}\BibitemShut {NoStop}%
\bibitem [{\citenamefont {Baumchen}\ and\ \citenamefont
  {Jacobs}(2010{\natexlab{a}})}]{Jacobs2010}%
  \BibitemOpen
  \bibfield  {author} {\bibinfo {author} {\bibfnamefont {O.}~\bibnamefont
  {Baumchen}}\ and\ \bibinfo {author} {\bibfnamefont {K.}~\bibnamefont
  {Jacobs}},\ }\bibfield  {title} {\enquote {\bibinfo {title} {Slip effects in
  polymer thin films},}\ }\href@noop {} {\bibfield  {journal} {\bibinfo
  {journal} {J. Phys. Cond. Mat.}\ }\textbf {\bibinfo {volume} {22}},\ \bibinfo
  {pages} {033102} (\bibinfo {year} {2010}{\natexlab{a}})}\BibitemShut
  {NoStop}%
\bibitem [{\citenamefont {Hatzikiriakos}(2012)}]{hatzikiriakos_wall_2012}%
  \BibitemOpen
  \bibfield  {author} {\bibinfo {author} {\bibfnamefont {Savvas~G.}\
  \bibnamefont {Hatzikiriakos}},\ }\bibfield  {title} {\enquote {\bibinfo {title} 
  {Wall slip in molten polymers},}\ }\href
  {\doibase 10.1016/j.progpolymsci.2011.09.004} {\bibfield  {journal} {\bibinfo
   {journal} {Progress in Polymer Science}\ }\textbf {\bibinfo {volume} {37}},\
  \bibinfo {pages} {624--643} (\bibinfo {year} {2012})}\BibitemShut {NoStop}%
\bibitem [{\citenamefont {Thompson}\ and\ \citenamefont
  {Robbins}(1990)}]{Thompsonrobbins90}%
  \BibitemOpen
  \bibfield  {author} {\bibinfo {author} {\bibfnamefont {P.A.}\ \bibnamefont
  {Thompson}}\ and\ \bibinfo {author} {\bibfnamefont {M.O.}\ \bibnamefont
  {Robbins}},\ }\bibfield  {title} {\enquote {\bibinfo {title} {Shear flow near
  solids: Epitaxial order and flow boundary conditions},}\ }\href@noop {}
  {\bibfield  {journal} {\bibinfo  {journal} {Phys. Rev. A}\ }\textbf {\bibinfo
  {volume} {41}},\ \bibinfo {pages} {6830} (\bibinfo {year}
  {1990})}\BibitemShut {NoStop}%
\bibitem [{\citenamefont {Bocquet}\ and\ \citenamefont
  {Barrat}(1994)}]{BarratBocquetPRE94}%
  \BibitemOpen
  \bibfield  {author} {\bibinfo {author} {\bibfnamefont {L.}~\bibnamefont
  {Bocquet}}\ and\ \bibinfo {author} {\bibfnamefont {J.-L.}\ \bibnamefont
  {Barrat}},\ }\bibfield  {title} {\enquote {\bibinfo {title} {Hydrodynamic
  boundary conditions, correlation functions, and kubo relations for confined
  fluids},}\ }\href@noop {} {\bibfield  {journal} {\bibinfo  {journal} {Phys.
  Rev. E}\ }\textbf {\bibinfo {volume} {49}},\ \bibinfo {pages} {3079--3092}
  (\bibinfo {year} {1994})}\BibitemShut {NoStop}%
\bibitem [{\citenamefont {Thompson}\ and\ \citenamefont
  {Troian}(1997)}]{thompsontroian97}%
  \BibitemOpen
  \bibfield  {author} {\bibinfo {author} {\bibfnamefont {P.A.}\ \bibnamefont
  {Thompson}}\ and\ \bibinfo {author} {\bibfnamefont {S.M.}\ \bibnamefont
  {Troian}},\ }\bibfield  {title} {\enquote {\bibinfo {title} {A general
  boundary condition for liquid flows at solid surfaces},}\ }\href@noop {}
  {\bibfield  {journal} {\bibinfo  {journal} {Nature}\ }\textbf {\bibinfo
  {volume} {389}},\ \bibinfo {pages} {360--362} (\bibinfo {year}
  {1997})}\BibitemShut {NoStop}%
\bibitem [{\citenamefont {Pit}\ \emph {et~al.}(2000)\citenamefont {Pit},
  \citenamefont {Hervet},\ and\ \citenamefont {L\'eger}}]{pitPRL2000}%
  \BibitemOpen
  \bibfield  {author} {\bibinfo {author} {\bibfnamefont {R.}~\bibnamefont
  {Pit}}, \bibinfo {author} {\bibfnamefont {H.}~\bibnamefont {Hervet}}, \ and\
  \bibinfo {author} {\bibfnamefont {L.}~\bibnamefont {L\'eger}},\ }\bibfield
  {title} {\enquote {\bibinfo {title} {Direct experimental evidence of slip in
  hexadecane : solid interfaces},}\ }\href@noop {} {\bibfield  {journal}
  {\bibinfo  {journal} {Phys. Rev. Lett.}\ }\textbf {\bibinfo {volume} {85}},\
  \bibinfo {pages} {980--983} (\bibinfo {year} {2000})}\BibitemShut {NoStop}%
\bibitem [{\citenamefont {Thomas}\ \emph {et~al.}(2014)\citenamefont {Thomas},
  \citenamefont {Charrault},\ and\ \citenamefont {Neto}}]{netoreview}%
  \BibitemOpen
  \bibfield  {author} {\bibinfo {author} {\bibfnamefont {L.}~\bibnamefont
  {Thomas}}, \bibinfo {author} {\bibfnamefont {E.}~\bibnamefont {Charrault}}, \
  and\ \bibinfo {author} {\bibfnamefont {C.}~\bibnamefont {Neto}},\ }\bibfield
  {title} {\enquote {\bibinfo {title} {Interfacial slip on rough, patterned and
  soft surfaces: A review of experiments and simulations},}\ }\href@noop {}
  {\bibfield  {journal} {\bibinfo  {journal} {Adv. in Coll. and Interf. Sci.}\
  }\textbf {\bibinfo {volume} {210}},\ \bibinfo {pages} {21--38} (\bibinfo
  {year} {2014})}\BibitemShut {NoStop}%
\bibitem [{\citenamefont {Secchi}\ \emph {et~al.}(2016)\citenamefont {Secchi},
  \citenamefont {Marbach}, \citenamefont {Nigu{\`e}s}, \citenamefont {Stein},
  \citenamefont {Siria},\ and\ \citenamefont
  {Bocquet}}]{BocquetSecchiNature2016}%
  \BibitemOpen
  \bibfield  {author} {\bibinfo {author} {\bibfnamefont {E.}~\bibnamefont
  {Secchi}}, \bibinfo {author} {\bibfnamefont {S.}~\bibnamefont {Marbach}},
  \bibinfo {author} {\bibfnamefont {A.}~\bibnamefont {Nigu{\`e}s}}, \bibinfo
  {author} {\bibfnamefont {D.}~\bibnamefont {Stein}}, \bibinfo {author}
  {\bibfnamefont {A.}~\bibnamefont {Siria}}, \ and\ \bibinfo {author}
  {\bibfnamefont {L.}~\bibnamefont {Bocquet}},\ }\bibfield  {title} {\enquote
  {\bibinfo {title} {Massive radius-dependent flow slippage in carbon
  nanotubes},}\ }\href@noop {} {\bibfield  {journal} {\bibinfo  {journal}
  {Nature}\ }\textbf {\bibinfo {volume} {537}} (\bibinfo {year}
  {2016})}\BibitemShut {NoStop}%
\bibitem [{\citenamefont {Leroy}\ \emph {et~al.}(2012)\citenamefont {Leroy},
  \citenamefont {Steinberger}, \citenamefont {Cottin-Bizonne}, \citenamefont
  {Restagno}, \citenamefont {L\'eger},\ and\ \citenamefont
  {Charlaix}}]{Leroy2012}%
  \BibitemOpen
  \bibfield  {author} {\bibinfo {author} {\bibfnamefont {S.}~\bibnamefont
  {Leroy}}, \bibinfo {author} {\bibfnamefont {A.}~\bibnamefont {Steinberger}},
  \bibinfo {author} {\bibfnamefont {C.}~\bibnamefont {Cottin-Bizonne}},
  \bibinfo {author} {\bibfnamefont {F.}~\bibnamefont {Restagno}}, \bibinfo
  {author} {\bibfnamefont {L.}~\bibnamefont {L\'eger}}, \ and\ \bibinfo
  {author} {\bibfnamefont {E.}~\bibnamefont {Charlaix}},\ }\bibfield  {title}
  {\enquote {\bibinfo {title} {Hydrodynamic interaction between a spherical
  particle and an elastic surface: A gentle probe for soft thin films},}\
  }\href@noop {} {\bibfield  {journal} {\bibinfo  {journal} {Phys. Rev. Lett.}\
  }\textbf {\bibinfo {volume} {108}},\ \bibinfo {pages} {264501} (\bibinfo
  {year} {2012})}\BibitemShut {NoStop}%
\bibitem [{\citenamefont {Baumchen}\ and\ \citenamefont
  {Jacobs}(2010{\natexlab{b}})}]{baumchen_slip_2010}%
  \BibitemOpen
  \bibfield  {author} {\bibinfo {author} {\bibfnamefont {O.}~\bibnamefont
  {Baumchen}}\ and\ \bibinfo {author} {\bibfnamefont {K.}~\bibnamefont
  {Jacobs}},\ }\bibfield  {title} {\enquote {\bibinfo {title} {Slip effects in
  polymer thin films},}\ }\href {<Go to ISI>://000273055400003} {\bibfield
  {journal} {\bibinfo  {journal} {Journal Of Physics-Condensed Matter}\
  }\textbf {\bibinfo {volume} {22}},\ \bibinfo {pages} {--} (\bibinfo {year}
  {2010}{\natexlab{b}})}\BibitemShut {NoStop}%
\bibitem [{\citenamefont {H\'enot}\ \emph {et~al.}(2017)\citenamefont
  {H\'enot}, \citenamefont {Chennevi\`ere}, \citenamefont {Drockenmuller},
  \citenamefont {L\'eger},\ and\ \citenamefont
  {Restagno}}]{henot_comparison_2017}%
  \BibitemOpen
  \bibfield  {author} {\bibinfo {author} {\bibfnamefont {Marceau}\ \bibnamefont
  {H\'enot}}, \bibinfo {author} {\bibfnamefont {Alexis}\ \bibnamefont
  {Chennevi\`ere}}, \bibinfo {author} {\bibfnamefont {Eric}\ \bibnamefont
  {Drockenmuller}}, \bibinfo {author} {\bibfnamefont {Liliane}\ \bibnamefont
  {L\'eger}}, \ and\ \bibinfo {author} {\bibfnamefont {Fr\'ed\'eric}\
  \bibnamefont {Restagno}},\ }\bibfield  {title} {\enquote {\bibinfo {title}
  {Comparison of the slip of a {PDMS} melt on weakly adsorbing surfaces
  measured by a new photobleaching-based technique},}\ }\href {\doibase
  10.1021/acs.macromol.7b00601} {\bibfield  {journal} {\bibinfo  {journal}
  {Macromolecules}\ ,\ \bibinfo {pages} {5592--5598}} (\bibinfo {year}
  {2017})}\BibitemShut {NoStop}%
\bibitem [{\citenamefont {Navier}(1823)}]{Navier}%
  \BibitemOpen
  \bibfield  {author} {\bibinfo {author} {\bibfnamefont {C.L.}\ \bibnamefont
  {Navier}},\ }\bibfield  {title} {\enquote {\bibinfo {title} {M{\'e}moires sur
  les lois du mouvement des fluides},}\ }\href@noop {} {\bibfield  {journal}
  {\bibinfo  {journal} {Memoires de l'Acad{\'e}mie des Sciences}\ }\textbf
  {\bibinfo {volume} {6}},\ \bibinfo {pages} {389--416} (\bibinfo {year}
  {1823})}\BibitemShut {NoStop}%
\bibitem [{\citenamefont {Dobrynin}\ and\ \citenamefont
  {Rubinstein}(2005)}]{DobryninRubinstein2005}%
  \BibitemOpen
  \bibfield  {author} {\bibinfo {author} {\bibfnamefont {A.V.}\ \bibnamefont
  {Dobrynin}}\ and\ \bibinfo {author} {\bibfnamefont {M.}~\bibnamefont
  {Rubinstein}},\ }\bibfield  {title} {\enquote {\bibinfo {title} {Theory of
  polyelectrolytes in solutions and at surfaces},}\ }\href@noop {} {\bibfield
  {journal} {\bibinfo  {journal} {Prog. in Polymer Science}\ }\textbf {\bibinfo
  {volume} {30}},\ \bibinfo {pages} {1049--1118} (\bibinfo {year}
  {2005})}\BibitemShut {NoStop}%
\bibitem [{\citenamefont {Chauveteau}\ \emph {et~al.}(1984)\citenamefont
  {Chauveteau}, \citenamefont {Tirrell},\ and\ \citenamefont
  {Omari}}]{Chauveteau1}%
  \BibitemOpen
  \bibfield  {author} {\bibinfo {author} {\bibfnamefont {Guy}\ \bibnamefont
  {Chauveteau}}, \bibinfo {author} {\bibfnamefont {M}~\bibnamefont {Tirrell}},
  \ and\ \bibinfo {author} {\bibfnamefont {A}~\bibnamefont {Omari}},\
  }\bibfield  {title} {\enquote {\bibinfo {title} {Concentration dependence of
  the effective viscosity of polymer solutions in small pores with repulsive or
  attractive walls},}\ }\href@noop {} {\bibfield  {journal} {\bibinfo
  {journal} {Journal of colloid and interface science}\ }\textbf {\bibinfo
  {volume} {100}},\ \bibinfo {pages} {41--54} (\bibinfo {year}
  {1984})}\BibitemShut {NoStop}%
\bibitem [{\citenamefont {Omari}\ \emph {et~al.}(1989)\citenamefont {Omari},
  \citenamefont {Moan},\ and\ \citenamefont {Chauveteau}}]{Chauveteau2}%
  \BibitemOpen
  \bibfield  {author} {\bibinfo {author} {\bibfnamefont {A}~\bibnamefont
  {Omari}}, \bibinfo {author} {\bibfnamefont {M.}~\bibnamefont {Moan}}, \ and\
  \bibinfo {author} {\bibfnamefont {G.}~\bibnamefont {Chauveteau}},\ }\bibfield
   {title} {\enquote {\bibinfo {title} {Wall effect in the flow of flexible
  polymer solution through small pores},}\ }\href@noop {} {\bibfield  {journal}
  {\bibinfo  {journal} {Rheolocica Acta}\ }\textbf {\bibinfo {volume} {28}},\
  \bibinfo {pages} {520--526} (\bibinfo {year} {1989})}\BibitemShut {NoStop}%
\bibitem [{\citenamefont {Cayer-Barrioz}\ \emph {et~al.}(2008)\citenamefont
  {Cayer-Barrioz}, \citenamefont {Mazuyer}, \citenamefont {Tonck},\ and\
  \citenamefont {Yamaguchi}}]{CayerBarriozTL2008}%
  \BibitemOpen
  \bibfield  {author} {\bibinfo {author} {\bibfnamefont {J.}~\bibnamefont
  {Cayer-Barrioz}}, \bibinfo {author} {\bibfnamefont {D.}~\bibnamefont
  {Mazuyer}}, \bibinfo {author} {\bibfnamefont {A.}~\bibnamefont {Tonck}}, \
  and\ \bibinfo {author} {\bibfnamefont {E.}~\bibnamefont {Yamaguchi}},\
  }\bibfield  {title} {\enquote {\bibinfo {title} {Drainage of a wetting
  liquid: Effective slippage or polymer depletion?}}\ }\href@noop {} {\bibfield
   {journal} {\bibinfo  {journal} {Tribo. Lett.}\ }\textbf {\bibinfo {volume}
  {32}},\ \bibinfo {pages} {81--90} (\bibinfo {year} {2008})}\BibitemShut
  {NoStop}%
\bibitem [{\citenamefont {Cuenca}\ and\ \citenamefont
  {Bodiguel}(2013)}]{CuencaPRL2013}%
  \BibitemOpen
  \bibfield  {author} {\bibinfo {author} {\bibfnamefont {A.}~\bibnamefont
  {Cuenca}}\ and\ \bibinfo {author} {\bibfnamefont {H.}~\bibnamefont
  {Bodiguel}},\ }\bibfield  {title} {\enquote {\bibinfo {title} {Submicron flow
  of polymer solutions: Slippage reduction due to confinement},}\ }\href@noop
  {} {\bibfield  {journal} {\bibinfo  {journal} {Phys. Rev. Lett.}\ }\textbf
  {\bibinfo {volume} {110}},\ \bibinfo {pages} {108304} (\bibinfo {year}
  {2013})}\BibitemShut {NoStop}%
\bibitem [{\citenamefont {De~Gennes}(1981)}]{de_gennes_polymer_1981}%
  \BibitemOpen
  \bibfield  {author} {\bibinfo {author} {\bibfnamefont {P.~G.}\ \bibnamefont
  {De~Gennes}},\ }\bibfield  {title} {\enquote {\bibinfo {title} {Polymer
  solutions near an interface. adsorption and depletion layers},}\ }\href
  {\doibase 10.1021/ma50007a007} {\bibfield  {journal} {\bibinfo  {journal}
  {Macromolecules}\ }\textbf {\bibinfo {volume} {14}},\ \bibinfo {pages}
  {1637--1644} (\bibinfo {year} {1981})}\BibitemShut {NoStop}%
\bibitem [{\citenamefont {Barnes}(1995{\natexlab{b}})}]{barnes1995review}%
  \BibitemOpen
  \bibfield  {author} {\bibinfo {author} {\bibfnamefont {Howard~A}\
  \bibnamefont {Barnes}},\ }\bibfield  {title} {\enquote {\bibinfo {title} {A
  review of the slip (wall depletion) of polymer solutions, emulsions and
  particle suspensions in viscometers: its cause, character, and cure},}\
  }\href@noop {} {\bibfield  {journal} {\bibinfo  {journal} {Journal of
  Non-Newtonian Fluid Mechanics}\ }\textbf {\bibinfo {volume} {56}},\ \bibinfo
  {pages} {221--251} (\bibinfo {year} {1995}{\natexlab{b}})}\BibitemShut
  {NoStop}%
\bibitem [{\citenamefont {Kuhl}\ \emph {et~al.}(1998)\citenamefont {Kuhl},
  \citenamefont {Berman}, \citenamefont {Hui},\ and\ \citenamefont
  {Israelachvili}}]{Kuhl-Israelachvili1998}%
  \BibitemOpen
  \bibfield  {author} {\bibinfo {author} {\bibfnamefont {T.L.}\ \bibnamefont
  {Kuhl}}, \bibinfo {author} {\bibfnamefont {A.D.}\ \bibnamefont {Berman}},
  \bibinfo {author} {\bibfnamefont {S.W.}\ \bibnamefont {Hui}}, \ and\ \bibinfo
  {author} {\bibfnamefont {J.N.}\ \bibnamefont {Israelachvili}},\ }\bibfield
  {title} {\enquote {\bibinfo {title} {Part 1. direct measurement of depletion
  attraction and thin film viscosity between lipid bilayers in aqueous
  polyethylene glycol solutions},}\ }\href@noop {} {\bibfield  {journal}
  {\bibinfo  {journal} {Macromolecules}\ }\textbf {\bibinfo {volume} {31}},\
  \bibinfo {pages} {8250--8257} (\bibinfo {year} {1998})}\BibitemShut {NoStop}%
\bibitem [{\citenamefont {Horn}\ \emph {et~al.}(2000)\citenamefont {Horn},
  \citenamefont {Vinogradova}, \citenamefont {Mackay},\ and\ \citenamefont
  {Phan-Thien}}]{HornVinogradovaJCP2000}%
  \BibitemOpen
  \bibfield  {author} {\bibinfo {author} {\bibfnamefont {R.G.}\ \bibnamefont
  {Horn}}, \bibinfo {author} {\bibfnamefont {O.I.}\ \bibnamefont
  {Vinogradova}}, \bibinfo {author} {\bibfnamefont {M.E.}\ \bibnamefont
  {Mackay}}, \ and\ \bibinfo {author} {\bibfnamefont {N.}~\bibnamefont
  {Phan-Thien}},\ }\bibfield  {title} {\enquote {\bibinfo {title} {Hydrodynamic
  slippage inferred from thin film drainage measurements in a solution of
  nonadsorbing polymer},}\ }\href@noop {} {\bibfield  {journal} {\bibinfo
  {journal} {J. Chem. Phys.}\ }\textbf {\bibinfo {volume} {112}} (\bibinfo
  {year} {2000})}\BibitemShut {NoStop}%
\bibitem [{\citenamefont {Knoben}\ \emph {et~al.}(2007)\citenamefont {Knoben},
  \citenamefont {Besseling},\ and\ \citenamefont
  {Cohen~Stuart}}]{knoben_direct_2007}%
  \BibitemOpen
  \bibfield  {author} {\bibinfo {author} {\bibfnamefont {W.}~\bibnamefont
  {Knoben}}, \bibinfo {author} {\bibfnamefont {N.~A.~M.}\ \bibnamefont
  {Besseling}}, \ and\ \bibinfo {author} {\bibfnamefont {M.~A.}\ \bibnamefont
  {Cohen~Stuart}},\ }\bibfield  {title} {\enquote {\bibinfo {title} {Direct
  measurement of depletion and hydrodynamic forces in solutions of a reversible
  supramolecular polymer},}\ }\href {\doibase 10.1021/la062656x} {\bibfield
  {journal} {\bibinfo  {journal} {Langmuir}\ }\textbf {\bibinfo {volume}
  {23}},\ \bibinfo {pages} {6095--6105} (\bibinfo {year} {2007})}\BibitemShut
  {NoStop}%
\bibitem [{\citenamefont {Restagno}\ \emph {et~al.}(2002)\citenamefont
  {Restagno}, \citenamefont {Crassous}, \citenamefont {Charlaix}, \citenamefont
  {Cottin-Bizonne},\ and\ \citenamefont {Monchanin}}]{rsi2002}%
  \BibitemOpen
  \bibfield  {author} {\bibinfo {author} {\bibfnamefont {F.}~\bibnamefont
  {Restagno}}, \bibinfo {author} {\bibfnamefont {J.}~\bibnamefont {Crassous}},
  \bibinfo {author} {\bibfnamefont {E.}~\bibnamefont {Charlaix}}, \bibinfo
  {author} {\bibfnamefont {C.}~\bibnamefont {Cottin-Bizonne}}, \ and\ \bibinfo
  {author} {\bibfnamefont {M.}~\bibnamefont {Monchanin}},\ }\bibfield  {title}
  {\enquote {\bibinfo {title} {A highly sensitive dynamic surface force
  apparatus for nanorheology},}\ }\href@noop {} {\bibfield  {journal} {\bibinfo
   {journal} {Rev. of Sci. Instr.}\ }\textbf {\bibinfo {volume} {73}},\
  \bibinfo {pages} {2292--2297} (\bibinfo {year} {2002})}\BibitemShut {NoStop}%
\bibitem [{\citenamefont {Garcia}\ \emph {et~al.}(2016)\citenamefont {Garcia},
  \citenamefont {Barraud}, \citenamefont {Picard}, \citenamefont {Giraud},
  \citenamefont {Charlaix},\ and\ \citenamefont {Cross}}]{Garcia2016}%
  \BibitemOpen
  \bibfield  {author} {\bibinfo {author} {\bibfnamefont {L.}~\bibnamefont
  {Garcia}}, \bibinfo {author} {\bibfnamefont {C.}~\bibnamefont {Barraud}},
  \bibinfo {author} {\bibfnamefont {C.}~\bibnamefont {Picard}}, \bibinfo
  {author} {\bibfnamefont {J.}~\bibnamefont {Giraud}}, \bibinfo {author}
  {\bibfnamefont {E.}~\bibnamefont {Charlaix}}, \ and\ \bibinfo {author}
  {\bibfnamefont {B.}~\bibnamefont {Cross}},\ }\bibfield  {title} {\enquote
  {\bibinfo {title} {A micro-nano-rheometer for the mechanics of soft matter at
  interfaces},}\ }\href@noop {} {\bibfield  {journal} {\bibinfo  {journal}
  {Rev. of Sci. Instr.}\ }\textbf {\bibinfo {volume} {87}},\ \bibinfo {pages}
  {113906} (\bibinfo {year} {2016})}\BibitemShut {NoStop}%
\bibitem [{\citenamefont {Pelletier}\ \emph {et~al.}(1994)\citenamefont
  {Pelletier}, \citenamefont {Montfort},\ and\ \citenamefont
  {Lapique}}]{Pelletier94b}%
  \BibitemOpen
  \bibfield  {author} {\bibinfo {author} {\bibfnamefont {E.}~\bibnamefont
  {Pelletier}}, \bibinfo {author} {\bibfnamefont {J.-P.}\ \bibnamefont
  {Montfort}}, \ and\ \bibinfo {author} {\bibfnamefont {F.}~\bibnamefont
  {Lapique}},\ }\bibfield  {title} {\enquote {\bibinfo {title} {Surface force
  apparatus and its application to nanorheological studies},}\ }\href@noop {}
  {\bibfield  {journal} {\bibinfo  {journal} {Journal of Rheology}\ }\textbf
  {\bibinfo {volume} {38}},\ \bibinfo {pages} {1151--68} (\bibinfo {year}
  {1994})}\BibitemShut {NoStop}%
\bibitem [{\citenamefont {Hocking}(1973)}]{Hocking1973}%
  \BibitemOpen
  \bibfield  {author} {\bibinfo {author} {\bibfnamefont {L.M.}\ \bibnamefont
  {Hocking}},\ }\bibfield  {title} {\enquote {\bibinfo {title} {The effect of
  slip on the motion of a sphere close to a wall and of two adjacent
  spheres},}\ }\href@noop {} {\bibfield  {journal} {\bibinfo  {journal} {J. of
  Engineering Mathematics}\ }\textbf {\bibinfo {volume} {7}},\ \bibinfo {pages}
  {207--221} (\bibinfo {year} {1973})}\BibitemShut {NoStop}%
\bibitem [{\citenamefont {Vinogradova}(1995)}]{vinogradova_drainage_1995}%
  \BibitemOpen
  \bibfield  {author} {\bibinfo {author} {\bibfnamefont {Olga~I.}\ \bibnamefont
  {Vinogradova}},\ }\bibfield  {title} {\enquote {\bibinfo {title} {Drainage of
  a thin liquid film confined between hydrophobic surfaces},}\ }\href {\doibase
  10.1021/la00006a059} {\bibfield  {journal} {\bibinfo  {journal} {Langmuir}\
  }\textbf {\bibinfo {volume} {11}},\ \bibinfo {pages} {2213--2220} (\bibinfo
  {year} {1995})}\BibitemShut {NoStop}%
\bibitem [{\citenamefont {Tuinier}\ and\ \citenamefont
  {Taniguchi}(2005)}]{tuinier_polymer_2005}%
  \BibitemOpen
  \bibfield  {author} {\bibinfo {author} {\bibfnamefont {Remco}\ \bibnamefont
  {Tuinier}}\ and\ \bibinfo {author} {\bibfnamefont {Takashi}\ \bibnamefont
  {Taniguchi}},\ }\bibfield  {title} {\enquote {\bibinfo {title} {Polymer
  depletion-induced slip near an interface},}\ }\href {\doibase
  10.1088/0953-8984/17/2/L01} {\bibfield  {journal} {\bibinfo  {journal}
  {Journal of Physics: Condensed Matter}\ }\textbf {\bibinfo {volume} {17}},\
  \bibinfo {pages} {L9--L14} (\bibinfo {year} {2005})}\BibitemShut {NoStop}%
\bibitem [{\citenamefont {Brenner}\ and\ \citenamefont
  {Howard}(2013)}]{Brenner2013}%
  \BibitemOpen
  \bibfield  {author} {\bibinfo {author} {\bibfnamefont {M.}~\bibnamefont
  {Brenner}}\ and\ \bibinfo {author} {\bibfnamefont {S.}~\bibnamefont
  {Howard}},\ }\href@noop {} {\emph {\bibinfo {title} {Interfacial transport
  processes and rheology}}}\ (\bibinfo  {publisher} {Elsevier},\ \bibinfo
  {year} {2013})\BibitemShut {NoStop}%
\end{thebibliography}
%

\textbf{\\Author Information} Correspondence and requests for materials should be addressed to 	Elisabeth.Charlaix@univ-grenoble-alpes.fr.

\end{document}